\documentclass[twocolumn,prb,amsmath,amssymb,showpacs]{revtex4}

\usepackage{graphicx}
\usepackage{bm}

\newcommand{\vect}[1]{\protect{\bm{\mathrm{#1}}}}
\newcommand{\matr}[1]{\protect{\bm{#1}}}
\newcommand{\phdag}{{\vphantom{\dagger}}}

\begin{document}
\title{Phase diagram and quasiparticle properties of the Hubbard model
within cluster two-site DMFT} 
\author{E. C. Carter}
\author{A. J. Schofield} 
\affiliation{
School of Physics and Astronomy, University of Birmingham, Edgbaston,
Birmingham B15 2TT, United Kingdom }%

\date{\today}

\begin{abstract}
We present a cluster dynamical mean-field treatment of the Hubbard
model on a square lattice to study the evolution of magnetism and
quasiparticle properties as the electron filling and interaction
strength are varied. Our approach for solving the dynamical mean-field
equations is an extension of Potthoff's ``two-site'' method
[Phys. Rev. B. {\bf 64}, 165114 (2001)] where the self-consistent bath
is represented by a highly restricted set of states. As well as the
expected antiferromagnetism close to half-filling, we observe
distortions of the Fermi surface. The proximity of a van Hove point
and the incipient antiferromagnetism lead to the evolution from an
electron-like Fermi surface away from the Mott transition, to a
hole-like one near half-filling. Our results also show a gap opening
anisotropically around the Fermi surface close to the Mott transition
(reminiscent of the pseudogap phenomenon seen in the cuprate
high-$T_c$ superconductors). This leaves Fermi arcs which are closed
into pockets by lines with very small quasiparticle residue.
\end{abstract}

\maketitle

\section{Introduction}

Understanding the physics of the Hubbard model continues to be a
fundamental issue in strongly correlated systems. This model captures
the transition between a metallic state and a correlated insulator;
how this transition takes place has been investigated by many workers
in the past, with differing approaches emphasizing the formation of
Hubbard bands,~\cite{hubbard_1964a} the increasing mass of the
quasiparticles~\cite{brinkman_1970a} and the proximity to
antiferromagnetism.~\cite{slater_1951a} Significant progress has been
made in recent years by the development of dynamical mean-field theory
(DMFT).~\cite{georges_1996a} Here a set of equations---exact in the
limit of infinite dimensions---is derived which maps the problem onto
an interacting impurity model to be solved self-consistently. This
method has revealed a Mott transition that is a synthesis of the
pictures of Mott and Brinkman--Rice: formation of Hubbard bands in
parallel with a large mass quasiparticle.  The absence of momentum
dependence in correlations within DMFT means that this method does not
account for variations in quasiparticle properties across the
Brillouin zone.  This is likely to be important if the Mott transition
has an antiferromagnetic aspect.

Building on the success of dynamical mean-field theory, extensions to
the theory are being actively studied. One such extension combines it
with density-functional theory to improve the treatment of local
correlations.~\cite{anisimov_1997a, lichtenstein_1998a,
lichtenstein_2001a} Another considers intersite correlations via so
called cluster DMFT.~\cite{hettler_1998a, kotliar_2001a,
lichtenstein_2000a} It is also often important to include multiple
local orbitals when modeling real materials such as the correlated
oxides. In all of these extensions the task of solving the resulting
self-consistent equations becomes increasingly problematic: the
equations require the solution of impurity models with an increasing
number of local degrees of freedom, which often demand
high-performance computing resources and sophisticated approximation
schemes to extract the low energy physics.

Yet in contrast to these computationally intensive approaches,
Potthoff~\cite{potthoff_2001a} demonstrated that much of the Mott
transition physics could be captured with a drastic approximation of
the non-interacting bath of electrons that couples to the impurity in
DMFT.  Whereas in one computational scheme for tackling the DMFT
equations, the bath is modeled by up to twelve coupled sites (the
exact diagonalization method~\cite{caffarel_1994a}), Potthoff
used just a single site to represent the bath.  The self-consistency
conditions were constructed to ensure that the quasiparticle
properties and band filling were matched.  Solving them yielded a
successful description of the Mott transition showing, for example,
the narrowing of the quasiparticle resonance and the formation of the
Hubbard bands.  A value of the critical Hubbard interaction, $U$, 
was obtained comparable to the best calculations.

In this paper we present an extension of Potthoff's ``two-site''
approach to treat cluster DMFT.  We use our method to investigate the
approach to the Mott transition for a single band Hubbard model on a
two-dimensional square lattice with nearest neighbor hopping, $t$. 
The Hamiltonian is
\begin{equation}
\hat{H}_{\rm Hub} = \sum_{{\vect k}, \sigma}
\epsilon^\phdag_{\vect k} \hat{c}^\dagger_{{\vect k},\sigma}
\hat{c}^{\phdag}_{{\vect k}, \sigma} + U \sum_i 
\hat{c}^\dagger_{i \uparrow}\hat{c}^{\phdag}_{i\uparrow}
\hat{c}^\dagger_{i \downarrow}\hat{c}^{\phdag}_{i\downarrow} \; ,
\label{HubbardHam}
\end{equation}
where $\epsilon_{\vect k} = -2t [\cos (k_xa) + \cos (k_ya)]$.  We
study this model as an example case, although our method is readily
extended to more complicated Hamiltonians. The method allows us to
quickly investigate the zero temperature phase diagram across a large
range of parameter space using a desktop computer. The efficacy of our
approach is reinforced by our results which are consistent with other
results in the literature for the magnetic phase diagram.  Moreover we
can go beyond existing work to study the quasiparticle properties in
momentum space: seeing, for example, how an electron-like metal (well
away from half-filling) becomes a hole-like doped Mott insulator near
half-filling. Some of our results are suggestive of the physics of the
cuprate superconductors with the appearance of pseudogap regions and
``arc-like'' Fermi surfaces brought about, in our case, by the
combination of antiferromagnetism and proximity to a van Hove point.
We also observe a Fermi surface distortion resulting from a
Pomeranchuk instability also reported elsewhere in the literature.%
\cite{halboth_2000a,hankevych_2002a,metzner_2003a,neumayr_2003a}
Physically, this is due to the proximity of the Fermi surface to the
van Hove point, but the tendency will be exaggerated in our model due
to the reduced symmetry of our cluster (see later).

We begin with a summary of DMFT together with Potthoff's ``two-site''
approach. We then discuss how DMFT can be extended to a cluster of
sites, and describe our application of Potthoff's approach to a
cluster consisting of a pair of sites. We then demonstrate this method
on the 2D Hubbard model and present our results. We conclude with a
discussion of the physics behind these results and future extensions
of our method.

\section{From DMFT to the two-site approach}
The DMFT procedure~\cite{georges_1996a} can be described as follows.
We focus on a single site of the Hubbard model, and notionally
integrate out all the other sites.  This gives an effective action for
the remaining site of the form
\begin{equation}\begin{split}
\label{dmftseff}
S_\text{eff} = &- \int_{0}^{\beta} d\tau
                   \int_{0}^{\beta} d\tau^{\prime}
c_{\sigma}^{\dagger}(\tau) 
\mathcal{G}^{-1}_0(\tau-\tau^{\prime})c_{\sigma}^{\phdag}(\tau^{\prime})\\
&+ U \int_{0}^{\beta} d\tau c^\dagger_{\uparrow}(\tau)c^\phdag_{\uparrow}(\tau)
c^\dagger_{\downarrow}(\tau)c^\phdag_{\downarrow}(\tau).
\end{split}\end{equation}
The function $\mathcal{G}_0$ completely encapsulates the dynamics of
electrons entering and leaving the site from the rest of the lattice;
however it is not known {\it a priori} since we cannot in practice
integrate out the other sites.  With this action we could determine
the interacting Green's function $G_{\text{local}}(i\omega_n)$, and
extract a local self-energy from the Dyson's equation
\begin{equation}
\label{dmftsigma}
G_{\text{local}}^{-1}(i\omega_n) = \mathcal{G}_0^{-1}(i\omega_n) -
\Sigma_\text{local}(i\omega_n).
\end{equation}
The DMFT ansatz, exact in infinite dimensions, is to use this local
self-energy as a spatially homogeneous (but frequency-dependent)
self-energy for the full lattice problem:
\begin{equation}
\label{dmftlatgf}
{G_{\text{lat}}(i\omega_n, \vect k)}^{-1} = i\omega_n + \mu -
\epsilon_{\vect k} - \Sigma_\text{local}(i\omega_n).
\end{equation}
The self-consistency requirement that the on-site Green's function of
the extended lattice (containing the self-energy) is the same as the
local Green's function we started with, {\it i.e.}
\begin{equation}
\label{dmftsc}
G_{\text{local}}(i\omega_n) = \sum_{\vect k} G_{\text{lat}}(i\omega_n,
\vect k),
\end{equation}
provides the constraint on the unknown initial function
$\mathcal{G}_0(i\omega_n)$, thereby completing the self-consistency
loop.

Eq.~\ref{dmftseff} is the effective action of a single interacting
impurity coupled to a continuum bath, but the procedure described
above cannot be achieved exactly because the single site impurity
problem with an unrestricted bath is still intractable. One must
approximate and use a model that is practically solvable; for example,
a finite-sized impurity model for exact
diagonalization,~\cite{caffarel_1994a} or a discretized effective
action for Quantum Monte Carlo methods.~\cite{jarrell_1992a} This
means that only limited functional forms of $\mathcal{G}_0(i\omega_n)$
can be represented, and also that the final self-consistency condition
cannot be implemented precisely.

In the ``two-site'' realization of DMFT introduced by
Potthoff,~\cite{potthoff_2001a} the local model is an impurity site
together with a bath consisting of {\em a single site only}, with a
Hamiltonian:
\begin{equation}
\label{impham}
\hat{H} = 
U \hat{a}^\dagger_{\uparrow} \hat{a}^\phdag_{\uparrow}
 \hat{a}^\dagger_{ \downarrow} \hat{a}^\phdag_{\downarrow} +
\sum_{\sigma}
\epsilon_{c\sigma} \hat{c}^\dagger_{\sigma} 
\hat{c}^\phdag_{\sigma} +V_\sigma (
\hat{a}^\dagger_{ \sigma} \hat{c}^\phdag_{\sigma} + \hat{c}^\dagger_{
 \sigma} \hat{a}^\phdag_{\sigma}) 
-\mu \hat{a}^\dagger_{\sigma}\hat{a}^\phdag_{\sigma},
\end{equation}
where the electron creation operators $\hat{a}^\dagger_{\sigma}$ and
$\hat{c}^\dagger_{\sigma}$ are for the impurity site and the bath site
respectively.  Diagonalizing the non-interacting ($U=0$) model yields
$\mathcal{G}_0(\omega)=\omega + \mu
-[V_\sigma^2/(\omega-\epsilon_{c\sigma})]$ (where we are now
considering zero temperature real frequency Green's functions).  The
two-site model allows a minimal frequency dependence in
$\mathcal{G}_0$.  By exactly diagonalizing the many particle
Hamiltonian of Eq.~\ref{impham}, the local on-impurity-site
interacting Green's function $G_{\text{imp}}(\omega)$ can be
constructed from the Lehmann representation, and the self-energy
$\Sigma(\omega)$ is extracted ({\it c.f.} Eq.~\ref{dmftsigma}) using:
\begin{equation}
\label{dmftsigmatwosite}
G_{\text{imp},\sigma}^{-1}(\omega) = \omega + \mu -
\frac{V_\sigma^2}{\omega-\epsilon_{c\sigma}} -
\Sigma_{\text{local},\sigma}(\omega).
\end{equation}

The full functional self-consistency of Eq.~\ref{dmftsc} cannot be
achieved within such a restricted representation so it is necessary to
decide how best to implement a self-consistency requirement.  Potthoff
chose two physically motivated features, taking advantage of the
analytic simplicity of the two-site impurity model.  Firstly, the
electron fillings given by the Green's functions must be equal for the
impurity model and the lattice model.  Secondly, features of the
central quasiparticle peak are matched: the self-energy is reduced to
the low energy form $\Sigma_{\text{local},\sigma}(\omega) \sim
a_\sigma + b_\sigma\omega$, and terms of the resulting ``coherent''
impurity and lattice Green's functions at high energy are matched (see
Ref.~\onlinecite{potthoff_2001a} for more detail).  In effect, the
shape of the central quasiparticle peak is analyzed by the size of its
high energy tails, isolated from other parts of the spectrum.  The
resulting self-consistency conditions for the four bath parameters
($V_\sigma, \epsilon_{c\sigma}$) are:
\begin{equation}
n_{\text{imp},\sigma} = n_{\text{lat},\sigma} \qquad \text{and} \qquad
V_\sigma^2 = \sum_{\vect k} \epsilon_{\vect k}^2 z_\sigma \; ,
\end{equation}
where the quasiparticle residue 
\begin{equation}
z_\sigma = \left( 1-
\frac{d\Sigma_{\text{local},\sigma}}{d\omega(0)}\right)^{-1} = 
\frac{1}{(1-b_\sigma)} \; , 
\end{equation}
and we have assumed that $\sum_{\vect k} \epsilon_{\vect k} = 0$.

Solving these equations produces results for the Mott transition which
compare well with the full DMFT, and properties of the Fermi liquid
which are consistent with exact results (see
Ref.~\onlinecite{potthoff_2001a}).  It is hard to imagine a simpler
model which can do this; it succeeds because of the physical
motivation of its self-consistency conditions, namely the filling and
properties of the quasiparticle peak.  It is a useful approach for
calculations on extended models such as multiple
bands~\cite{koga_2004a} and, as we shall demonstrate, clusters.

\section{From cluster DMFT to two-site pair-cluster DMFT}
We now consider an extension of this method to cluster
DMFT:~\cite{hettler_1998a, kotliar_2001a} instead of starting with a
single site, self-consistency conditions are derived for a cluster of
sites, which allows a momentum dependence in the self-energy.  The
geometry of the lattice becomes important and hence different types of
magnetic order can be investigated, and spectral information varies
non-trivially in $\vect k$-space, unlike conventional DMFT.  A number
of studies have been
reported.~\cite{lichtenstein_2000a,huscroft_2001a,maier_2002a,dahnken_2003a,parcollet_2003a}
We describe here a cluster DMFT for the case where the cluster
consists of just a pair of sites.

We then detail how to implement Potthoff's two-site method to solve
cluster DMFT.  This can be contrasted with the exact diagonalization
formulation of conventional DMFT,~\cite{caffarel_1994a} where a single
impurity site with multiple bath sites is used; instead, we use
multiple impurity sites in a cluster and include correlations in the
simplest way via Potthoff's scheme for single bath sites. Below, we
derive the self-consistency conditions for the pair-cluster (each of
the two cluster sites is connected to a single bath site).

Implementing a cluster DMFT is fundamentally ambiguous, as recently
noted by Biroli {\it et al.};~\cite{biroli_2003a} the approach we adopt is
``CDMFT'' within their classification, which is arguably the simplest
scheme appropriate for broken symmetry states.  One of the strengths
of conventional DMFT is that it is exact in the limit of infinite
dimensions. In contrast, our cluster approach is not a systematic
$1/d$ correction, though it is of course exact in the limit of
infinite cluster size when the self-consistent bath is merely a
sophisticated boundary condition.~\cite{hettler_1998a}

In cluster DMFT we imagine integrating out all sites except those in
the cluster. An electron can in general now leave from and arrive back
at any of the sites within the cluster: the function $\mathcal{G}_0$
must become a matrix, coupling together the dynamics of the cluster
sites.  The resulting action ({\it c.f.}  Eq.~\ref{dmftseff}) is:
\begin{equation}\begin{split}
\label{dmftclusterseff}
S_\text{eff} = &-\int_{0}^{\beta} \! d\tau \int_{0}^{\beta}
d\tau^{\prime} \sum_{i,j \in \{A,B\}} c_{i\sigma}^{\dagger}(\tau)
\mathcal{G}^{-1}_{0,ij}
(\tau-\tau^{\prime})c_{j\sigma}^{\phdag}(\tau^{\prime})\\
&+ U \int_{0}^{\beta} d\tau \sum_{i \in \{A,B\}}
c^\dagger_{i\uparrow}(\tau)c^\phdag_{i\uparrow}(\tau)
c^\dagger_{i\downarrow}(\tau)c^\phdag_{i\downarrow}(\tau) \; ,
\end{split}\end{equation}
where the summations are over the cluster sites.  Solving this local
problem now yields a matrix $G_{\text{local},ij}$ and a matrix
self-energy.

Different approaches to cluster DMFT involve different ways of
combining the matrix self-energy with the non-interacting lattice
Green's function, and different self-consistency conditions.  We shall
now describe our particular approach, for the specific case of a 2D
square lattice with a cluster consisting of a pair of sites. The sites
represent the two sublattices of the bipartite square lattice, and we
label them $A$ and $B$.  The impurity model consists of a pair of
sites connected by a $-t$ hopping element, cut out of the original
lattice [see Fig.~\ref{impfig}(a)].  Each site has its own independent
bath site [Fig.~\ref{impfig}(b)].  The Hamiltonian for this
pair-cluster impurity model is diagonalized as previously, resulting
in a $2\times 2$ matrix Green's function, whose self-energy matrix is
extracted ({\it c.f.}  Eq.~\ref{dmftsigmatwosite}) using Dyson's
equation:
\begin{widetext}
\begin{equation}
\matr{G}_{\text{imp},\sigma}^{-1}(\omega) = 
\begin{pmatrix} \omega+\mu - 
\frac{V_{A\sigma}^2}{\omega-\epsilon_{cA,\sigma}} 
-\Sigma_{A\sigma}(\omega) & t-\Sigma_{AB,\sigma}(\omega) \\
t-\Sigma_{AB,\sigma}(\omega)  & 
\omega+\mu - \frac{V_{B\sigma}^2}{\omega-\epsilon_{cB,\sigma}} 
-\Sigma_{B\sigma}(\omega)  \end{pmatrix} \; ,
\end{equation}
where the matrix index is the cluster site $(A,B)$.

\begin{figure*}
\includegraphics[width=0.7\textwidth]{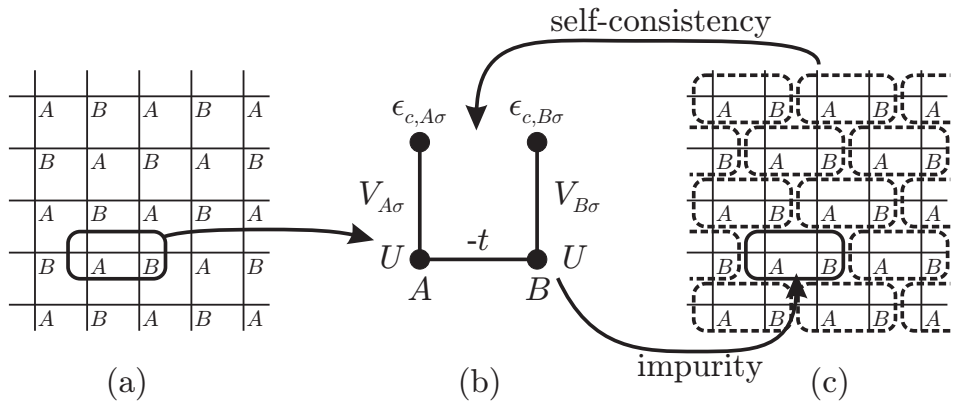}
\caption{\label{impfig} In our pair-cluster approach we identify a
small cluster in the lattice (a), and treat this cluster as an
``impurity'' connected to a bath.  A highly truncated basis is chosen
for this bath, leading to an impurity model with four sites (b).  To
reconstruct the extended system, the cluster is re-embedded in the
lattice periodically (c).  The bath is then adjusted to ensure it is
self-consistent with this embedding.  Although our cluster shape
manifestly breaks the $x$-$y$ symmetry, there are self-consistent
solutions which reflect the original square lattice symmetry.  }
\end{figure*}

The non-interacting lattice Green's function now has $\vect
k$-dependence from the Fourier transform of hopping elements to
surrounding clusters [Fig.~\ref{impfig}(c)], and is a matrix with
respect to the cluster sites $(A,B)$.  Combining it with the matrix
self-energy from the impurity model according to the DMFT ansatz,
gives the interacting lattice Green's function (as above, {\it c.f.}
Eq.~\ref{dmftlatgf}):
\begin{equation}
\label{latgf}
\matr{G}_{\text{lat},\sigma}^{-1}(\omega,\vect k) = 
\begin{pmatrix} \omega+\mu -\Sigma_{A\sigma}(\omega) & 
-\epsilon_{\vect k} e^{i k_x a}-\Sigma_{AB,\sigma}(\omega) \\
-\epsilon_{\vect k} e^{-i k_x a}-\Sigma_{AB,\sigma}(\omega) &
\omega+\mu -\Sigma_{B\sigma}(\omega) \end{pmatrix}.
\end{equation}\end{widetext}
Full self-consistency would require that the matrix version of
Eq.~\ref{dmftsc} be satisfied, but here because of the restricted
representation of $\mathcal{G}_{0,ij}$ using just two additional
non-interacting sites, we must adopt a more modest
self-consistency requirement.  We generalize the approach of
Ref.~\onlinecite{potthoff_2001a} and require firstly that the filling
of each sublattice is the same for the impurity and lattice models:
\begin{equation}
\label{clustersc1}
n_{\text{imp},\{A,B\},\sigma} = n_{\text{lat},\{A,B\},\sigma} \; ,
\end{equation}
which will effectively constrain $\epsilon_{c,\{A,B\},\sigma}$.  The
second condition characterizes the coherent quasiparticle
peak. Expanding the Green's function matrices at high $\omega$, with
the coherent form of the self-energy in place, gives a matrix equation
self-consistency equation involving a quasiparticle residue matrix
$\matr{Z} = (\matr{I} - d\matr{\Sigma}/d\omega(0))^{-1}$:
\begin{equation}\begin{split}
&\begin{pmatrix} \mu & t \\ t & \mu \end{pmatrix} \matr{Z}
\begin{pmatrix} \mu & t \\ t & \mu \end{pmatrix}
+\begin{pmatrix}  V^2_{A\sigma} & 0 \\ 0 & V^2_{B\sigma} \end{pmatrix} \\
= &
\sum_{\vect k} \begin{pmatrix} \mu &  
-\epsilon_{\vect k} e^{i k_x a} \\  
-\epsilon_{\vect k} e^{i k_x a} & 
\mu \end{pmatrix} \matr{Z}
\begin{pmatrix} \mu 
&  -\epsilon_{\vect k} e^{i k_x a} \\  
-\epsilon_{\vect k} e^{i k_x a} & \mu \end{pmatrix} \; .
\end{split}\end{equation}

After the $\vect k$-sum has been carried out we obtain the
self-consistency condition:
\begin{equation}
\label{clustersc2}
\begin{pmatrix}  V^2_{A\sigma} & 0 \\ 0 & V^2_{B\sigma} \end{pmatrix} =
\begin{pmatrix} 3t^2 Z_{B\sigma} & 0 \\ 0 &  3t^2 Z_{A\sigma} 
\end{pmatrix} \; .
\end{equation}
This constrains $\{V_A,V_B\}$ in terms of the quasiparticle residues
$\matr{Z}$ calculated from the local self-energy.

\section{Calculations}
We now describe the implementation of our cluster DMFT approach.  The
eight self-consistent equations
(Eqs~\ref{clustersc1} and~\ref{clustersc2}) are solved for the eight bath
parameters $\epsilon_{c,\{A,B\},\sigma}, V_{\{A,B\},\sigma}$, and
additionally we wish to consider a given total filling $n$, and this
provides a ninth condition which constrains the chemical potential
$\mu$ such that $\sum n_{\{A,B\},\sigma} = n$.  The practical problem
is thus nine-dimensional root finding, for which we use a Broyden method
combined with line searches.~\cite{press_1992a}

The two main computations are diagonalizing the Hamiltonian and
calculating the band filling for the lattice case.  For the latter, we
sum the imaginary part of the lattice Green's function
(Eq.~\ref{latgf}) over a finite number of points in $\vect k$-space
and then integrate numerically for $\omega: (-\infty,0)$.  A small
analytic continuation $\omega - i \delta$ is required, with $\delta
\sim 0.005 t$, and typically the $\vect k$-sum involves $K=120$ points
across the Brillouin zone; checks were done to ensure that dependence
on $\delta$ and $K$ is insignificant.  Fillings above $n=0.98$ are
ignored since the $\vect k$-space resolution is insufficient to give
an accurate representation of the Fermi surface, except at exactly
half-filling where this is not an issue.

For each choice of $(U,n)$ there is in general more than one set of
bath parameters which satisfies the self-consistency conditions.
Phases with paramagnetic, ferromagnetic, antiferromagnetic,
ferrimagnetic, insulating and charge-ordered character have all
emerged. We did not consider superconducting order.  A pitfall of any
self-consistency scheme is the production of unphysical excited
states, and an advantage of the two-site DMFT is the ease of
calculating the energy of a solution.  Using Fetter and
Walecka,~\cite{fetter_1971a} the energy of a fully interacting system
is
\begin{equation}
\int_{-\infty}^{0} \hspace{-1em}d\omega \sum_{\vect k} 
\frac{1}{\pi} \underset{\omega\rightarrow \omega-i\delta}{\text{Im}} 
\text{tr} \left[ \left(\matr{H}_0(\omega,\vect k) +
\frac{1}{2}\matr{\Sigma}(\omega)\right) \cdot
\matr{G}_{\text{lat}}(\omega,\vect k) \right], 
\end{equation}
where $\matr{H}_0$ (the non-interacting Hamiltonian),
$\matr{\Sigma}$ and $\matr{G}_{\text{lat}}$ are matrices with respect
to $(A,B)$ as above. We use this to identify the lowest energy solution.

As well as the energy and magnetic order, we wish to access
the $\vect k$-space spectral information of our solutions.  Combining
sublattice Green's functions defined in the reduced Brillouin zone
into a single correlation function where $\vect k$ is in the extended
(non-symmetry-broken) Brillouin zone, gives:
\begin{widetext}
\begin{equation}
\label{qpgf}
G_{\vect k} = 
\frac{\omega + \mu - (\Sigma_A(\omega)+\Sigma_B(\omega))/2 +\epsilon_{\vect k}
+\Sigma_{AB}(\omega) \cos k_x a }%
{(\omega + \mu - \Sigma_A(\omega))(\omega + \mu -
  \Sigma_B(\omega))-\Sigma_{AB}^2(\omega) - 2 \Sigma_{AB}(\omega)
  \epsilon_{\vect k} \cos k_x a - \epsilon_{\vect k}^2 } \; .
\end{equation}\end{widetext}

\section{Results}

\begin{figure}
\includegraphics[width=\columnwidth]{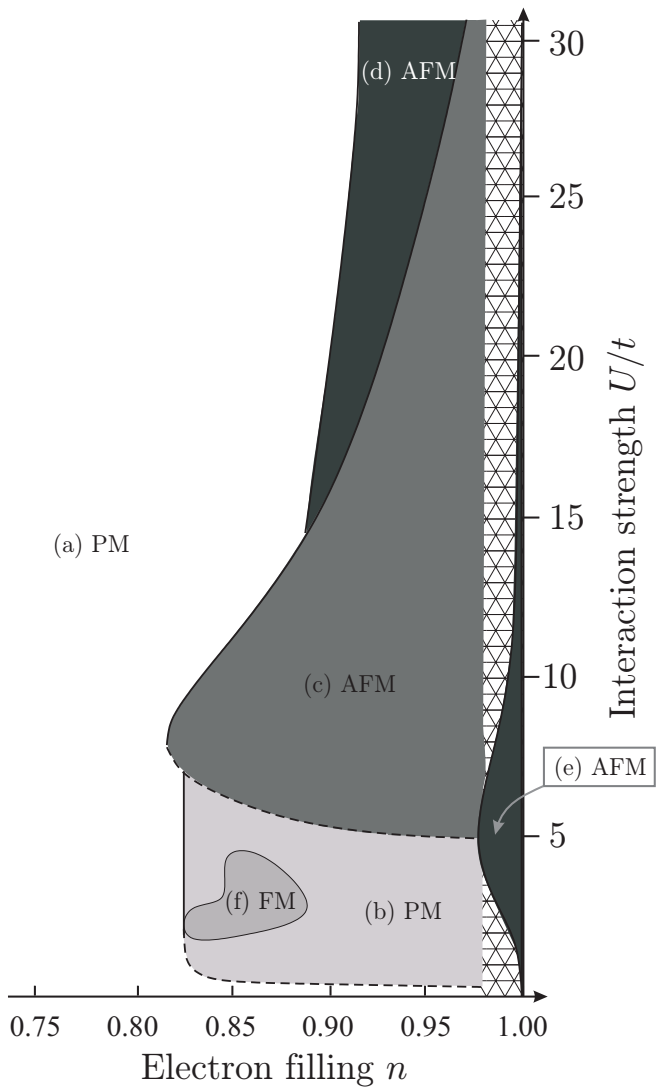}
\caption{\label{pd1} Schematic ground state phase diagram $(n,U/t)$
for the square lattice Hubbard model (bandwidth $8t$) for our cluster
DMFT, showing a conventional paramagnetic phase (a), a paramagnetic
phase with a distorted Fermi surface (b), a Slater-like
antiferromagnetic phase (c), another antiferromagnetic phase (d), a
pure Slater nested antiferromagnetic phase (e) and a ferromagnetic
phase (f).  A dashed line indicates a second order transition. The
phases are characterized further in the text.  Solutions for
the hatched region, $0.98<n<1$, are not considered (see text).}
\end{figure}

We apply our cluster DMFT approach to the single band Hubbard model on
a 2D square lattice defined in Eq.~\ref{HubbardHam}.  We analyze our
results for the region $0 \leq U/t \leq 30$ and filling $0.75 \leq n
\leq 1$. We find a ground state phase diagram with a variety of
phases, both magnetic and non-magnetic, as illustrated in
Fig.~\ref{pd1}; the extent of magnetism is much reduced from that
predicted by Hartree--Fock theory.  Note that all our phases are
metallic for $n\ne 1$ and insulating for the half-filled case $n=1$.
From an analysis of the full spectral function $A(\omega,\vect k)$ we
can characterize the phases and offer a physical origin for their
stability.

Recall that for this model there are three important factors driving
the underlying physics: the Mott transition, the nesting of the $U=0$
Fermi surface at $n=1$, and the van Hove points in the free particle
dispersion at $(\pi,0)$ and $(0,\pi)$.  Clearly at $n=1$ all of these
factors play a r\^ole, but away from this point we see the three
factors exerting distinct influences on the ground state physics.

\subsection{Trends across the phase diagram}

\begin{figure}
\includegraphics[width=\columnwidth]{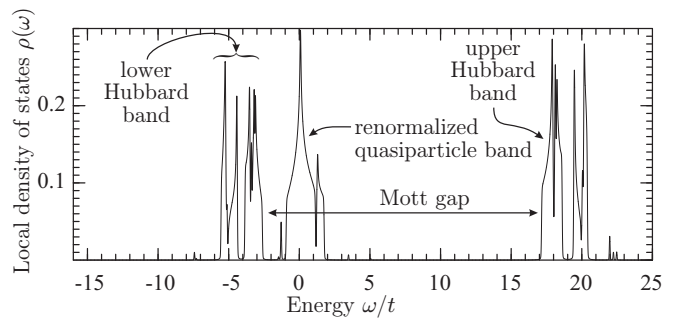}
\caption{
\label{dos1}
At $U=20t$, $n=0.8$ the ground state is paramagnetic [Fig.~\ref{pd1}
(a)]. The smoothed local density of states for this solution is
shown.  The central quasiparticle peak shows a single-particle density
of states, with a bandwidth $8t^* \sim 3t$ significantly renormalized
from the original $8t$ due to correlations.  Above and below the peak
there are upper and lower Hubbard bands, separated by a Mott gap of
size $\sim U$, and both Hubbard bands show imprints of the
single-particle dispersion.}
\end{figure}

The solution far from half-filling ($n<0.82$), or for very small
interaction strength ($U\lesssim 0.25t$), is a metallic paramagnetic
phase [Fig.~\ref{pd1} phase (a)] with a sharply defined Fermi surface,
seen as a discontinuity in $n_{\vect k}$ [Fig.~\ref{ghost1}(b), dotted
line].  This Fermi liquid becomes increasingly correlated as $U$ is
increased; the quasiparticle residue $z$ decreases, and $n_{\vect k}$
becomes more homogeneous (though maintaining a discontinuity) as weight
spreads out through $\vect k$-space. This reflects the local nature of
the strong repulsion, $U$.  Spectrally, Hubbard side bands are
observed to appear, and the central quasiparticle band narrows; an
example density of states is shown in Fig.~\ref{dos1}, consistent with
the three peak structure~\footnote{The quasiparticle peak is
  disconnected from the upper and lower Hubbard bands, which is a
  feature of the Potthoff approach~\cite{potthoff_2001a}} used to
understand the Mott transition.~\cite{georges_1996a}

\begin{figure}
\begin{center}
\includegraphics[width=0.8\columnwidth]{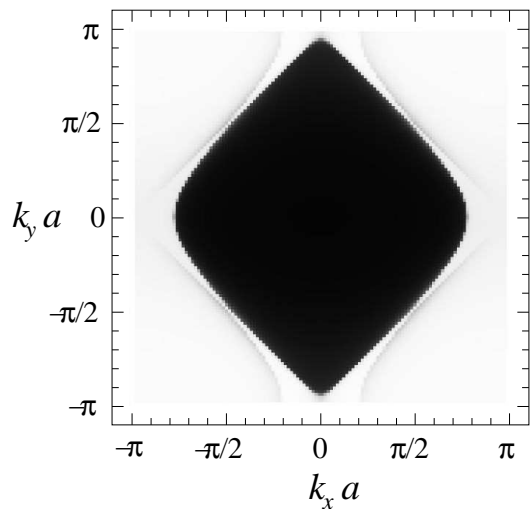}
\end{center}
\caption{
\label{fs1} 
At $U=2t$, $n=0.90$ the ground state is paramagnetic but the Fermi
surface has spontaneously distorted [Fig.~\ref{pd1} phase (b)]. Here
the electron filling $n_{\vect k}$ is shown in grayscale across the
Brillouin zone: black corresponds to fully occupied states, white are
completely empty states.  The Fermi surface is well defined, but has
undergone a Pomeranchuk transition: electrons have moved from
$(\pi,0)$ to $(0,\pi)$, exploiting the flat dispersion near the van
Hove point. Note the appearance of a faint ``ghost'' Fermi surface
displaced from the original by a wavevector $(\pi/a, \pi/a)$. This is
due to the effect of antiferromagnetic fluctuations.}
\end{figure}

As the filling $n$ is increased nearer to 1 (half-filling), the Fermi
surface in the paramagnetic state expands, and we observe a phase
transition [to phase (b) of Fig.~\ref{pd1}] where the Fermi surface
breaks $x$-$y$ symmetry (see Fig.~\ref{fs1}).  It is likely that the
system is exploiting the reduced Fermi velocity near the van Hove
point $(0,\pi)$ to redistribute electrons in momentum space to lower
the total energy---a Pomeranchuk transition. This tendency has
also been seen by other
authors.~\cite{halboth_2000a,hankevych_2002a,metzner_2003a,neumayr_2003a}
Note however, our cluster breaks $x$-$y$ symmetry from the outset, so
this is not strictly symmetry breaking, but the degree of deviation
from $x$-$y$ symmetry increases abruptly at this point.

\begin{figure}
\includegraphics[width=\columnwidth]{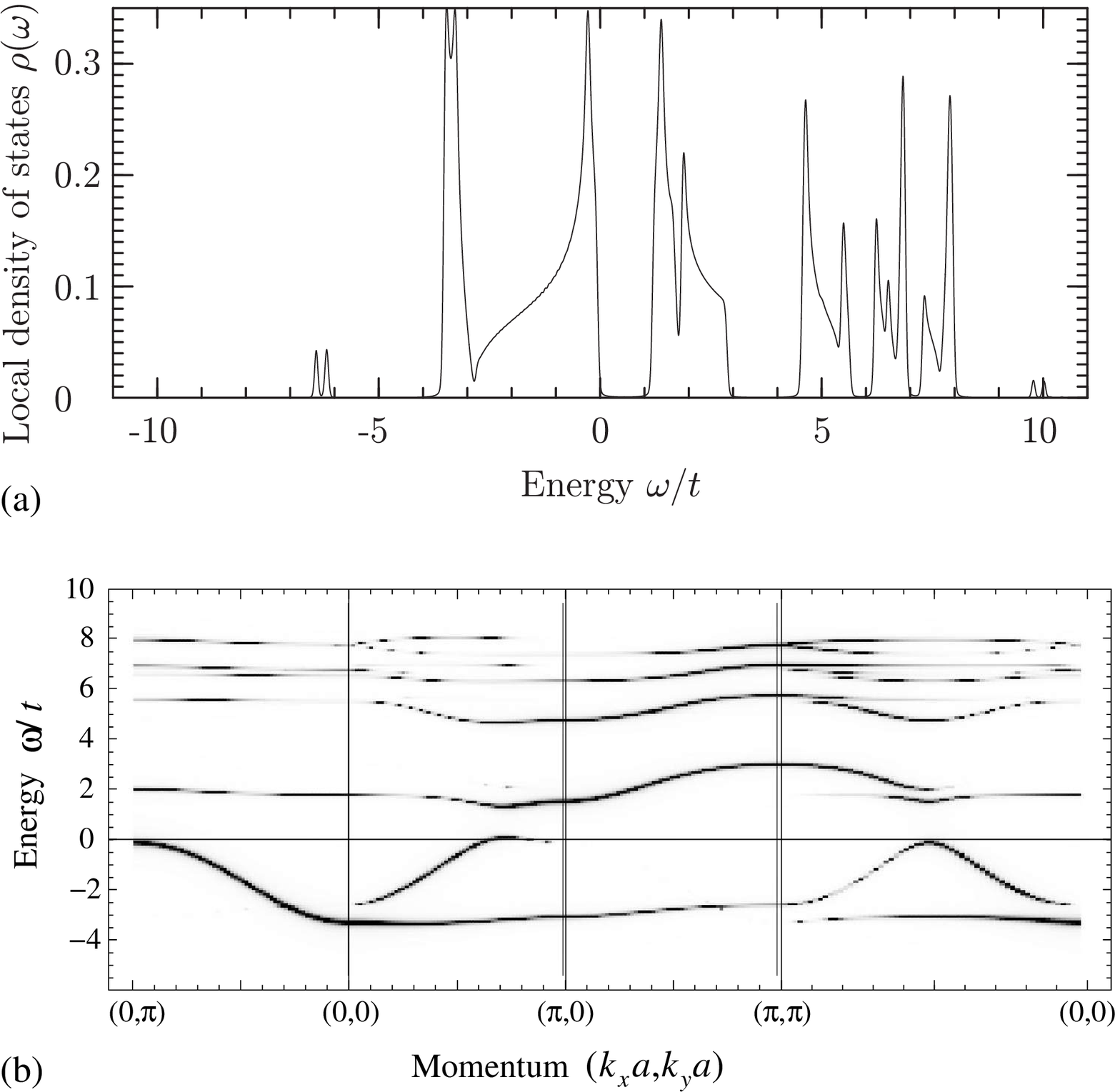}
\caption{
\label{dos2} 
At $U=6t$, $n=0.88$ the ground state is an antiferromagnetic metal
with a distorted Fermi surface [Fig.~\ref{pd1} phase (c)]. (a) 
The total local density of states is shown for this state 
(smoothed and combining up and down spins).
The magnetic gap is apparent, and correlations cause
further partitioning. In this region the gap size increases with
increasing $U$.  The densities of states for the neighboring
paramagnetic solution [Fig.~\ref{pd1} phase (b)] are insignificantly
different, although the sublattice- and spin-resolved densities of
states are of course unbalanced for antiferromagnetic
solutions.
(b) The evolution of the electron spectral function is shown for
various straight-line sections through the Brillouin zone. The plot is
grayscale with darker regions representing states with higher spectral
weight. As well as weakly dispersing bands at high energies, the
quasiparticle dispersion can be seen for energies between $\pm 4t$. 
The form of the dispersion is that expected for a mean field
spin-density wave state. Hence we identify the gap in the density of
states as being Slater-like.}
\end{figure}

For $U\gtrsim 6t$ and $n \gtrsim 0.8$, we find that the ground state
is a spin-symmetry-broken antiferromagnetic phase [Fig.~\ref{pd1}
phase (c)].  The Fermi surface is Pomeranchuk distorted, remaining the
same shape as for the paramagnet [Fig.~\ref{pd1} phase (b)].  The
local density of states shown in Fig.~\ref{dos2}(a) changes little
between that paramagnetic and this magnetically ordered state.
Indeed, the only change is that $(n_{A\uparrow}=n_{B\downarrow})$
becomes no longer equal to $(n_{A\downarrow}=n_{B\uparrow})$.  We
interpret this phase as a Slater antiferromagnet because the density
of states and quasiparticle dispersion is, for low $U$ at least, what
one would expect from a weak spin-density wave transition on a
metallic state with a gap growing like $U$ (see, for example, the
mean-field theory of Schrieffer, Wen and
Zhang~\cite{schrieffer_1989a}).  Note that here the gap in the density
of states is already present in the paramagnetic state; we will refer
back to this point in relation to the pseudogap physics of the
cuprates.

\begin{figure}
\includegraphics[width=\columnwidth]{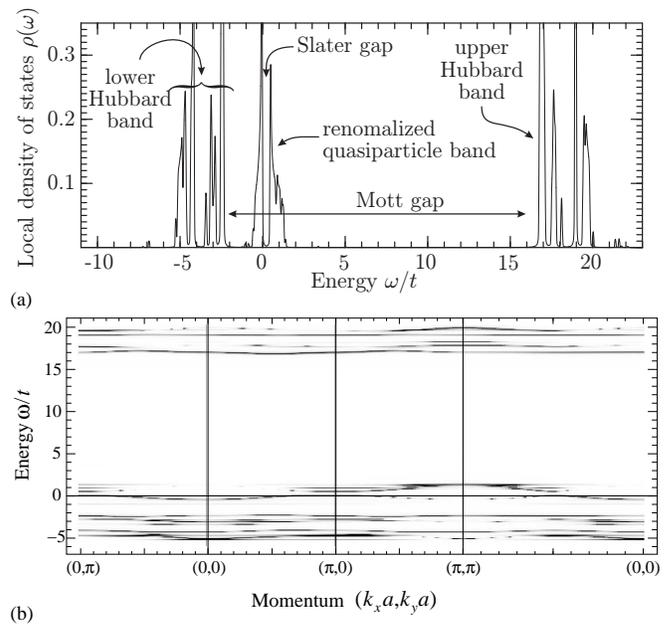}
\caption{
\label{dos3}
At $U=20t$, $n=0.9$ the ground state is also an antiferromagnetic
metal with little $x-y$ symmetry breaking [Fig.~\ref{pd1} phase (d)]. 
(a) Here we show the
local density of states (smoothed, up and down
spins combined).  The central
quasiparticle peak is separated from upper and lower Hubbard bands and
is heavily renormalized due to correlations. It is also
gapped; this gap decreases with increasing $U$.
(b) From the spectral density we identify the dispersing quasiparticle
near the chemical potential. The form of the dispersion is very
similar to that of Fig.~\ref{dos2}(b), so Slater-like, but the degree of 
dispersion is very much reduced.}
\end{figure}

There is a second antiferromagnetic phase [Fig.~\ref{pd1} phase (d)]
which appears when $U\gtrsim 16t$.  Here the magnetic gap develops
within the narrow quasiparticle peak of the density of states, which
is itself well separated from two fully formed Hubbard bands in the
familiar three peak structure (see Fig.~\ref{dos3}). The momentum
dependence of the narrow quasiparticle dispersion has the same
structural form as that in the Slater magnetic phase above but with a
greatly reduced bandwidth. This can be seen by comparing the spectral
functions shown in Fig.~\ref{dos2}(b) and Fig.~\ref{dos3}(b) near
$\omega=0$. The magnetism is thus appearing like a spin-density wave
transition but now for the renormalized quasiparticles. We find that
the gap reduces as $U$ is increased consistent with $\sim t^2/U$; thus
we interpret this phase as an antiferromagnetic metal in an effective
$t$-$J$ model, where $t_{\rm eff} \ll t$ and $J_{\rm eff} \sim t^2/U$.
The neighboring paramagnetic phase (at a reduced filling of $n
\lesssim 0.9$) could perhaps be described, at low energies, as a
nearly-antiferromagnetic Fermi liquid.

There are two other magnetic phases.  First, exactly at half-filling
(and extending to lower $n$ near $U\sim 5t$) there is a
antiferromagnet [Fig.~\ref{pd1} phase (e)] of fairly pure Slater
character, inevitable for a bipartite lattice.  The spectrum we find
is consistent with that described elsewhere in the literature, for
example the separate high energy bands in
Ref.~\onlinecite{dahnken_2003a}.  Second, there is a small patch of
low moment ferromagnetism [Fig.~\ref{pd1} phase (f)]; here the system
is exploiting the soft Fermi surface near the van Hove points ({\it
c.f.}  Ref.~\onlinecite{metzner_2003a}), and the Fermi surface for one
spin species distorts more than the other.

\subsection{Metal--insulator transition and the pseudogap}

A key puzzle in the study of correlated electrons near the Mott
transition concerns the changes in the Fermi surface as the Mott
insulating state is approached.  How, for example, does the large
volume Fermi surface expected in a Fermi liquid obeying Luttinger's
theorem, evolve into a doped Mott insulator where it appears that the
number of holes characterizes physical properties (see, for example,
Ref.~\onlinecite{uemura_1989a})? This question can be answered within
the constraints of our approach: as the carrier concentration
approaches half-filling, we see a large Fermi surface breaking up into
hole pockets separated by regions without free carriers.  In this
section we will describe this process in more detail.

\begin{figure*}
\begin{center}
\includegraphics[clip,width=0.8\textwidth]{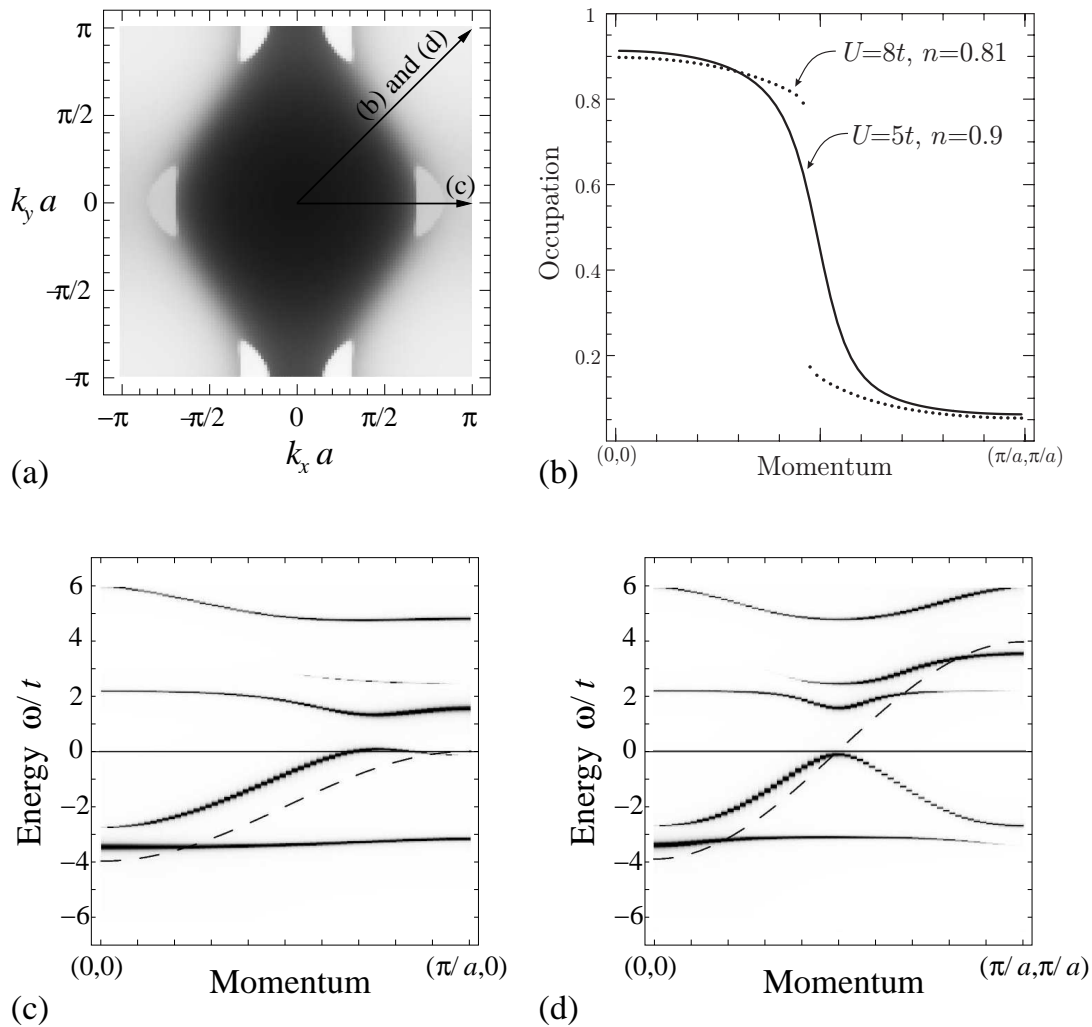}
\end{center}
\caption{
\label{ghost1}
At $U=5t$, $n=0.9$ the ground state is paramagnetic and breaks $x$-$y$
symmetry. It also shows the breakup of the Fermi surface into hole
pockets by the formation of an anisotropic gap around the Fermi
surface. 
Fig.~(a) shows $n_{\vect k}$ in grayscale across the
Brillouin zone (black $\leftrightarrow 1$, white $\leftrightarrow
0$). A distinct Fermi surface can be seen along some directions but
not others.  
Fig.~(b) illustrates the absence of a Fermi surface
discontinuity in $n_{\mathbf k}$ along the line $(0,0)
\rightarrow (\pi,\pi)$ (solid line).  This should be compared to the
form of $n_{\mathbf k}$ in a phase without such a gap [$(U=8t,
n=0.81)$ shown by the dotted line]. The nature of the anisotropy
around the Fermi surface can be seen in the electron spectral functions: 
(c) A grayscale plot of the spectral function (dark $\leftrightarrow$
larger weight) along the line $(0,0) \rightarrow (\pi,0)$. 
The chemical potential $\mu$ crosses a modified dispersion
twice, creating the hole pocket---though with little weight at the
second crossing point.
(d) The spectral function (dark $\leftrightarrow$ larger weight) along
the line $(0,0) \rightarrow (\pi,\pi)$ shows that $\mu$ falls within
a gap; there is no Fermi surface along this line.  The dashed
line in (c) and (d) shows the non-interacting
dispersion. Fig.~\ref{ghostsch} shows schematically the origin of this
electronic structure. 
}
\end{figure*}

Well away from half-filling, we observe a distinct Fermi surface,
defined by a discontinuity in $n_\vect k$ [see dotted line in
Fig.~\ref{ghost1}(b)], with a Luttinger volume equal to $n$.  As
$n\rightarrow 1$ and we approach an antiferromagnetic state, evidence
of a ``ghost'' Fermi surface is seen: low energy electron-like
excitations appear at the position of the original Fermi surface but
displaced by the antiferromagnetic wavevector $(\pi,\pi)$. This can be
seen as a faint line in Fig.~\ref{fs1}.  Note that the magnetic
symmetry has not yet been broken.

\begin{figure}
\includegraphics[clip,width=\columnwidth]{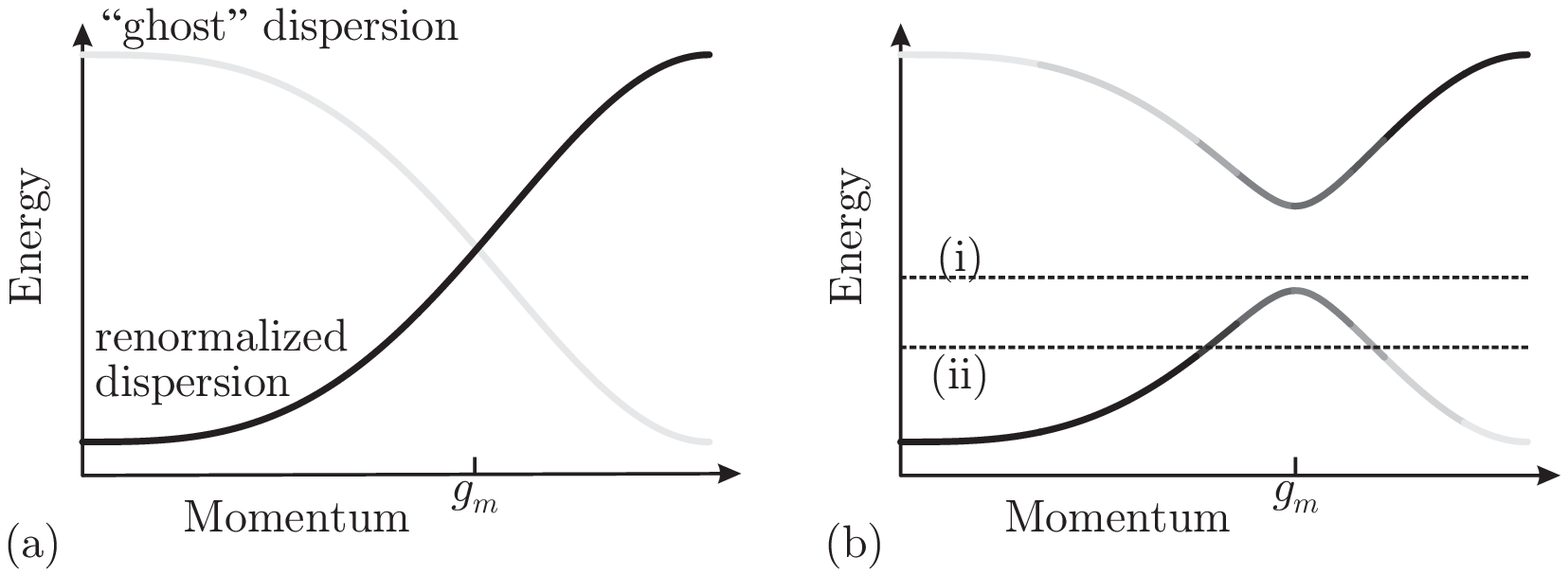}
\caption{
\label{ghostsch}
A schematic view of the origin of the structure of the electron
spectral function in the pseudogap region [as seen within our approach
in Figs~\ref{ghost1}(c) and~(d)]. Darker
lines carry more spectral weight. 
(a) Distinct fluctuations on the $A$
and $B$ sublattices result in a second dispersion of low weight (a
``ghost'' dispersion). This is formed by the folding of the original, 
possibly renormalized, dispersion in the antiferromagnetic Brillouin
zone (defining $g_m$).
(b) This hybridizes with the original dispersion and forms a gap. 
Along directions in momentum space where the
chemical potential lies in the gap [line(i)], there is a gap in the
excitation spectrum and no Fermi surface. Along directions where 
the chemical potential lies away from the gap it crosses the
dispersion at two points
[line (ii)]. A hole pocket forms with low spectral weight at the
second crossing.}
\end{figure}

Looking at the complete spectrum [for example in Figs~\ref{ghost1}(c)
and~(d)] reveals that this ``ghost'' Fermi surface is a manifestation
of a ``ghost dispersion''---the original dispersion shifted by
$(\pi,\pi)$.  This ghost dispersion has much less spectral weight than
the original renormalized quasiparticle dispersion.  Nevertheless, it
hybridizes with the original dispersion to give the final electronic
structure.  The hybridization opens a gap in the dispersion, and the
effects of the gap depend on whether the chemical potential $\mu$
falls within the gap, and how the gap evolves round the Brillouin
zone. A schematic view of the origin of the electronic structure is
given in Fig.~\ref{ghostsch}.

In the case of Fig.~\ref{fs1} where $U=2t$ and $n=0.90$, $\mu$ is
below the gap everywhere on the Fermi surface, and crosses the new
dispersion in two places; the spectral weight at the second crossing
is much weaker and causes the ``ghost'' Fermi surface. However, if the
filling is increased (raising $\mu$) or the interaction $U$ is
increased, widening the hybridization gap, then the chemical potential
may fall within the gap at various places around the Brillouin
zone. Such an occurrence is illustrated in Fig.~\ref{ghost1}(d) for
the case ($U=5t$, $n=0.90$) along the line in the Brillouin zone (0,0)
to ($\pi,\pi$). With no band crossing the Fermi surface there is no
discontinuity in $n_k$ as demonstrated in Fig.~\ref{ghost1}(b).
Instead $n_k$ falls smoothly because of the reduced spectral weight in
the ``ghost'' band that forms at the high momentum end of the low energy
dispersion. There is a gap for single electron excitations along this
direction.

This gap is a pseudogap rather than a complete gap as there are other
directions where a well-defined Fermi-surface remains.
Fig~\ref{ghost1}(c) shows the spectral function along the line (0,0)
to ($\pi$,0). There the chemical potential crosses the spectral
function dispersion in two places. The spectral weight is much weaker
at the higher momentum crossing because this part of the dispersion
derives mainly from the ``ghost'' band. These two crossings mark the
edges of a hole pocket as can be seen in Fig.~\ref{ghost1}(a). The
quasiparticles have developed hole-like characteristics and the
Luttinger volume has changed. However, because of the evolving
spectral weight in momentum space, only one side of the pocket has
quasiparticles with high electron weight. These hole pockets could
resemble disconnected Fermi surface arcs to experimental probes unable
to resolve the small electron weight at the second dispersion
crossing. A similar breakup of the Fermi surface near the Mott
transition was suggested by Furakawa {\it et
al.}~\cite{furukawa_1998a}

As $n$ increases, the hole pockets reduce in size until they become
points, and subsequently vanish and the system is an insulator.  This
illustrates how a smooth metal-insulator transition is possible.
After a pseudogap has formed, the Luttinger volume cannot be defined
as usual; the active Fermi surface consists of hole pockets, and rest
of the filled region of $\vect k$-space is bounded by portions of the
original Fermi surface where $n_{\vect k}$ falls smoothly, without a
step.

The features described above are reminiscent of the cuprate
high-temperature superconductors in their normal state, which are
believed to have a Fermi surface consisting of arc-like segments
separated by pseudogap regions (see for example
Refs~\onlinecite{chubukov_1997a,damascelli_2003a,yoshida_2003a}).  A
picture where the dispersion relation hybridizes with a ``ghost''
dispersion is consistent with angle-resolved photoemission
spectroscopy (ARPES) experiments (see for example Fig.~58 of
Ref.~\onlinecite{damascelli_2003a}), and also emerges from other
theoretical models.\cite{chubukov_1997a}

There are important differences, however, between our results and what
is seen in the cuprates, most significantly in the position of the
arcs and the pseudogap regions.  In the cuprates the Fermi surface
arcs are seen at $(\pi/2,\pi/2)$ but appear near $(0.7\pi,0)$ in our
calculations, and the pseudogap occurs near $(\pi,0)$ in the cuprates
but near $(\pi/2,\pi/2)$ in our calculations.  However, our results
are not incompatible, as restriction to a cluster with only a pair of
sites severely constrains the momentum dependence of the electron
self-energy.  Fermi surface shapes can only be formed from the
original dispersion $\epsilon_{\vect k}$ in combination with the
asymmetric dispersion $\epsilon_{\vect k} \cos k_xa$ (as Eq.~\ref{qpgf}
shows).  This does not admit the $d$-wave symmetry observed in the
cuprates but only $s$- and $p$-wave configurations, which means for
example that arcing can only happen near the van Hove points such as
$(\pi,0)$.  Adding more sites to the cluster would progressively optimize
our approximation to the real shape, and we intend to extend our work
to a four site ($2\times 2$) cluster which would permit $d$-wave
symmetry, though at the price of greater numerical complexity.

\section{Conclusion}
We have presented the zero temperature properties of a cluster-type
extension of dynamical mean-field theory, where the self-consistency
equations have been solved in a very restricted basis.  Our approach
allows us to calculate the full phase diagram of a fully interacting
system, and it can be surveyed and understood through access to
complete spectral information.  The basis we chose is ``two-site
DMFT'', whose results compare favorably with other theories and
experiment; our cluster extension additionally allows a limited
momentum dependence of the self-energy.

Our results provide clues to several puzzling features of the
metal--insulator transition and how antiferromagnetism, proximity to
van Hove points, and formation of the Mott gap compete with one
another; all play a r\^ole in various regimes.  The van Hove point
causes Fermi surface distortions even at low $U$; nesting causes
Slater antiferromagnetism near half-filling; and the Mott gap becomes
significant at high $U$ as one might have anticipated. In addition, we
see how $t$-$J$ model physics manifests itself at large $U$ via the small
magnetic gap.  We also see how, as the metal--insulator transition is
approached, the Fermi surface evolves from a renormalized Fermi liquid
obeying Luttinger's theorem, to a pseudogap state where a gap opens on
some parts of the Fermi surface breaking it up into hole pockets (with
a strongly momentum-dependent spectral density).

Our work is easily extensible to more sophisticated cluster DMFT
schemes or multiple bands, and we intend to study a $2\times 2$
cluster which allows tetragonal symmetry and would compare more
directly with the cuprates.

We thank C. Hooley, J. Quintanilla and Q. Si for helpful
discussions.  We gratefully acknowledge the support of the Royal
Society and the Leverhulme Trust (AJS) and EPSRC (ECC).


\begin{thebibliography}{10}
\expandafter\ifx\csname bibnamefont\endcsname\relax
  \def\bibnamefont#1{#1}\fi
\expandafter\ifx\csname bibfnamefont\endcsname\relax
  \def\bibfnamefont#1{#1}\fi
\expandafter\ifx\csname url\endcsname\relax
  \def\url#1{\texttt{#1}}\fi
\expandafter\ifx\csname urlprefix\endcsname\relax\def\urlprefix{URL }\fi
\providecommand{\eprint}[2][]{\url{#2}}

\bibitem{hubbard_1964a}
\bibfnamefont{J.}~\bibnamefont{Hubbard}.
\newblock Proc. R. Soc. London Ser. A \textbf{281}, 401 (1964).

\bibitem{brinkman_1970a}
\bibfnamefont{W.~F.} \bibnamefont{Brinkman} \bibnamefont{and}
  \bibfnamefont{T.~M.} \bibnamefont{Rice}.
\newblock Phys. Rev. B \textbf{2}, 4302 (1970).

\bibitem{slater_1951a}
\bibfnamefont{J.~C.} \bibnamefont{Slater}.
\newblock Phys. Rev. \textbf{82}, 538 (1951).

\bibitem{georges_1996a}
\bibfnamefont{A.}~\bibnamefont{Georges}, \emph{et~al.}
\newblock Rev. Mod. Phys. \textbf{68}, 13 (1996).

\bibitem{anisimov_1997a}
\bibfnamefont{V.~I.} \bibnamefont{Anisimov}, \emph{et~al.}
\newblock J. Phys. Cond. Mat. \textbf{9}, 7359 (1997).

\bibitem{lichtenstein_1998a}
\bibfnamefont{A.~I.} \bibnamefont{Lichtenstein} \bibnamefont{and}
  \bibfnamefont{M.~I.} \bibnamefont{Katsnelson}.
\newblock Phys. Rev. B \textbf{57}, 6884 (1998).

\bibitem{lichtenstein_2001a}
\bibfnamefont{A.~I.} \bibnamefont{Lichtenstein}, \bibfnamefont{M.~I.}
  \bibnamefont{Katsnelson}, \bibnamefont{and}
  \bibfnamefont{G.}~\bibnamefont{Kotliar}.
\newblock Phys. Rev. Lett. \textbf{87}, 067205 (2001).

\bibitem{hettler_1998a}
\bibfnamefont{M.~H.} \bibnamefont{Hettler}, \emph{et~al.}
\newblock Phys. Rev. B \textbf{58}, R7475 (1998).

\bibitem{kotliar_2001a}
\bibfnamefont{G.}~\bibnamefont{Kotliar}, \emph{et~al.}
\newblock Phys. Rev. Lett. \textbf{87}, 186401 (2001).

\bibitem{lichtenstein_2000a}
\bibfnamefont{A.~I.} \bibnamefont{Lichtenstein} \bibnamefont{and}
  \bibfnamefont{M.~I.} \bibnamefont{Katsnelson}.
\newblock Phys. Rev. B \textbf{62}, R9283 (2000).

\bibitem{potthoff_2001a}
\bibfnamefont{M.}~\bibnamefont{Potthoff}.
\newblock Phys. Rev. B \textbf{64}, 165114 (2001).

\bibitem{caffarel_1994a}
\bibfnamefont{M.}~\bibnamefont{Caffarel} \bibnamefont{and}
  \bibfnamefont{W.}~\bibnamefont{Krauth}.
\newblock Phys. Rev. Lett. \textbf{72}, 1545 (1994).

\bibitem{halboth_2000a}
\bibfnamefont{C.~J.} \bibnamefont{Halboth} \bibnamefont{and}
  \bibfnamefont{W.}~\bibnamefont{Metzner}.
\newblock Phys. Rev. Lett. \textbf{85}, 5162 (2000).

\bibitem{hankevych_2002a}
\bibfnamefont{V.}~\bibnamefont{Hankevych},
  \bibfnamefont{I.}~\bibnamefont{Grote}, \bibnamefont{and}
  \bibfnamefont{F.}~\bibnamefont{Wegner}.
\newblock Phys. Rev. B \textbf{66}, 094516 (2002).

\bibitem{metzner_2003a}
\bibfnamefont{W.}~\bibnamefont{Metzner}, \bibfnamefont{D.}~\bibnamefont{Rohe},
  \bibnamefont{and} \bibfnamefont{S.}~\bibnamefont{Andergassen}.
\newblock Phys. Rev. Lett. \textbf{91}, 066402 (2003).

\bibitem{neumayr_2003a}
\bibfnamefont{A.}~\bibnamefont{Neumayr} \bibnamefont{and}
  \bibfnamefont{W.}~\bibnamefont{Metzner}.
\newblock Phys. Rev. B \textbf{67}, 035112 (2003).

\bibitem{jarrell_1992a}
\bibfnamefont{M.}~\bibnamefont{Jarrell}.
\newblock Phys. Rev. Lett. \textbf{69}, 168 (1992).

\bibitem{koga_2004a}
\bibfnamefont{A.}~\bibnamefont{Koga}, \emph{et~al.}
\newblock \eprint{cond-mat/0401223}.

\bibitem{maier_2002a}
\bibfnamefont{T.~A.} \bibnamefont{Maier},
  \bibfnamefont{T.}~\bibnamefont{Pruschke}, \bibnamefont{and}
  \bibfnamefont{M.}~\bibnamefont{Jarrell}.
\newblock Phys. Rev. B \textbf{66}, 075102 (2002).

\bibitem{huscroft_2001a}
\bibfnamefont{C.}~\bibnamefont{Huscroft}, \emph{et~al.}
\newblock Phys. Rev. Lett. \textbf{86}, 139 (2001).

\bibitem{dahnken_2003a}
\bibfnamefont{C.}~\bibnamefont{Dahnken}, \emph{et~al.}
\newblock \eprint{cond-mat/0309407}.

\bibitem{parcollet_2003a}
\bibfnamefont{O.}~\bibnamefont{Parcollet},
  \bibfnamefont{G.}~\bibnamefont{Biroli}, \bibnamefont{and}
  \bibfnamefont{G.}~\bibnamefont{Kotliar}.
\newblock \eprint{cond-mat/0308577}.

\bibitem{biroli_2003a}
\bibfnamefont{G.}~\bibnamefont{Biroli},
  \bibfnamefont{O.}~\bibnamefont{Parcollet}, \bibnamefont{and}
  \bibfnamefont{G.}~\bibnamefont{Kotliar}.
\newblock \eprint{cond-mat/0307587}.

\bibitem{press_1992a}
\bibfnamefont{W.~H.} \bibnamefont{Press}, \emph{et~al.}
\newblock \emph{Numerical Recipes in C} (Cambridge University Press, 1992).

\bibitem{fetter_1971a}
\bibfnamefont{A.~L.} \bibnamefont{Fetter} \bibnamefont{and}
  \bibfnamefont{J.~D.} \bibnamefont{Walecka}.
\newblock \emph{Quantum Theory of Many-Particle Systems} (McGraw-Hill, 1971).
\newblock Eqn (9.36).

\bibitem{schrieffer_1989a}
\bibfnamefont{J.~R.} \bibnamefont{Schrieffer}, \bibfnamefont{X.~G.}
  \bibnamefont{Wen}, \bibnamefont{and} \bibfnamefont{S.~C.}
  \bibnamefont{Zhang}.
\newblock Phys. Rev. B \textbf{39}, 11663 (1989).

\bibitem{uemura_1989a}
\bibfnamefont{Y.~J.} \bibnamefont{Uemura}, \emph{et~al.}
\newblock Phys. Rev. Lett. \textbf{62}, 2317 (1989).

\bibitem{furukawa_1998a}
\bibfnamefont{N.}~\bibnamefont{Furukawa}, \bibfnamefont{T.~M.}
  \bibnamefont{Rice}, \bibnamefont{and}
  \bibfnamefont{M.}~\bibnamefont{Salmhofer}.
\newblock Phys. Rev. Lett. \textbf{81}, 3195 (1998).

\bibitem{chubukov_1997a}
\bibfnamefont{A.~V.} \bibnamefont{Chubukov} \bibnamefont{and}
  \bibfnamefont{D.~K.} \bibnamefont{Morr}.
\newblock Phys. Reports \textbf{288}, 355 (1997).

\bibitem{damascelli_2003a}
\bibfnamefont{A.}~\bibnamefont{Damascelli},
  \bibfnamefont{Z.}~\bibnamefont{Hussain}, \bibnamefont{and}
  \bibfnamefont{Z.-X.} \bibnamefont{Shen}.
\newblock Rev. Mod. Phys. \textbf{75}, 473 (2003).

\bibitem{yoshida_2003a}
\bibfnamefont{T.}~\bibnamefont{Yoshida}, \emph{et~al.}
\newblock Phys. Rev. Lett. \textbf{91}, 027001 (2003).

\end{thebibliography}

\end{document}